\documentclass{phostproc}
\usepackage{amsmath,bm,graphicx,color,titlesec,cuted} 

\newcommand{\Ro}{\mathrm{Ro}}
\newcommand{\rmv}{\mathrm{v}}

\newcommand{\Rowt}{\mathrm{\widetilde{Ro}_w}}
\newcommand{\Row}{\mathrm{Ro_w}}
\newcommand{\Roc}{\mathrm{Ro_c}}
\newcommand{\rmsub}{\mathrm{sub}}
\newcommand{\rmsup}{\mathrm{sup}}

\shorttitle{Rotating Convection: Gravito-Inertial Waves}
\shortauthors{Augustson \& Mathis}

\title{Rotating Convection and Gravito-Inertial Wave Generation in Stellar Interiors} 

\author{K.~C. Augustson$^1$, S. Mathis$^1$}
\affiliation{$^1$AIM, CEA, CNRS, Universit\'{e} Paris-Saclay, Universit\'{e} Paris Diderot, Sarbonne Paris Cit\'{e}, F-91191 Gif-sur-Yvette Cedex, France}

\abs{Gravito-inertial waves can be excited at the interface of convective and radiative regions and by the Reynolds
  stresses in the bulk of the convection zone. The magnitude of their energy flux will therefore vary with the
  properties of the convection. To assess how convection changes with rotation, a simplified local monomodal model for
  rotating convection is presented that provides the magnitude of the rms velocity, degree of superadiabaticity, and
  characteristic length scale as a function of the convective Rossby number as well as with thermal and viscous
  diffusivities. In the context of this convection model, two models for assessing the gravito-inertial wave flux are
  considered: an interfacial model and a full treatment of the Reynolds stress impact on the waves. It is found that
  there are regimes where the sub-inertial waves may carry a significant energy flux relative to pure gravity waves that
  depend upon the convective Rossby number, the ratio of the buoyancy time-scale in the stable region to the convective
  overturning time, and the wave frequency.}

\begin{document}

\maketitle

\section{Introduction}\label{sec:intro}

Convection in stars is driven through buoyancy and doubly-diffusive instabilities, where it primarily serves to
transport the energy released deep within the star or planet through regions where radiative energy transport is
inefficient.  Moreover, convection directly and nonlocally transports heat and chemicals through advection, entrainment,
and dissipative processes \citep[e.g.,][]{miesch09,kupka17,garaud18}.  In the presence of rotation, convection can also
transport angular momentum through Reynolds stresses and meridional flows to establish and maintain a differential
rotation \citep[e.g.,][]{glatzmaier82,kichatinov86,brun02,miesch09}. Convective motions that penetrate into subadiabatic
regions also generate waves that propagate in the otherwise quiescent regions, providing another means for transport and
mixing in those regions \citep[e.g.,][]{zahn97,kumar99,mathis09}. Gravito-inertial waves can be generated both near the
transition between convectively stable and unstable regions and throughout the convection zone itself, while they
propagate into the radiation zone \citep[e.g.,][]{press81,rogers13,lecoanet15}. They therefore transport angular
momentum and chemical species, which are deposited where these waves are dissipated either through thermal diffusion,
nonlinear wave breaking \citep[e.g.,][]{rogers15,rogers17}, or at critical layers \citep{alvan13,jouve14}.  Such waves
have been identified through asteroseismology as one of the candidates to explain angular momentum transport within
stellar radiation zones across the entire Hertzsprung-Russell diagram \citep[e.g.,][]{garcia07,saio15,aerts17}.  Indeed,
it has been demonstrated that for the Sun, evolved low-mass stars, and early-type stars stochastically excited internal
gravity waves can lead to an efficient transport of angular momentum \citep[e.g.,][]{talon05,mathis13a,pincon17}. When
rotation are present, gravito-inertial waves \citep[e.g.,][]{dintrans99,dintrans00,davidson13}, as well as new
instabilities, can be excited.  They can lead to other channels of global-scale transport phenomena, some of which can
become important through the coupling of inertial waves to convection throughout the convectively unstable zone
\citep[e.g.,][]{mathis13a,mathis14,fuller15b,stello16,lecoanet17}. Therefore, the impact of rotation on both the driving
mechanism of internal waves in stars and the transport induced by those waves needs to be understood and clarified.

In this work, a generalized version of a heuristic model for convection for rotating systems is presented following
\citet{stevenson79}, \citet{barker14}, and \citet{augustson18}. This model of convection is then employed to estimate
the gravito-inertial wave flux into the stable region adjacent to the convective region. Following the arguments of
\citet{press81} and \citet{andre17}, the interfacial generation of gravito-inertial waves and their associated energy
flux is assessed with respect to this convection model in \S\ref{sec:waves}. Subsequently, an estimate is given for the
gravito-inertial wave flux for waves excited by Reynolds stresses using the wave amplitudes derived in \citet{mathis14}
and the turbulence model constructed here and in \citet{lecoanet13}. A summary of the results and perspectives are
presented in \S\ref{sec:final}.

\section{General Framework}\label{sec:genframe}

The heuristic model will be considered to be local such that the length scales of the flow are much smaller than either
density or pressure scale heights. This is equivalent to ignoring the global dynamics and assuming that the convection
can be approximated as local at each radius and colatitude in a star or planet.  As such, one may consider the dynamics
to be Boussinesq. In other words, the model consists of an infinite Cartesian plane of a nearly incompressible fluid
with a small thermal expansion coefficient $\alpha_T=-\partial\ln{\rho}/\partial T|_P$ that is confined between two
impenetrable plates differing in temperature by $\Delta T$ and separated by a distance $\ell_0$. As seen in many papers
regarding Boussinesq dynamics \citep[e.g.,][]{chandrasekhar61}, the linearized Boussinesq equations can be reduced to a
single third-order in time and eighth-order in space equation for the vertical velocity.  The difference here, and in
the work of \citet{stevenson79}, is that the state that the system is being linearized about is nonlinearly saturated,
meaning that the potential temperature gradient is given by the Malkus-Howard convection theory
\citep[e.g.,][]{malkus54,howard63}.  Together, these equations provide a dispersion relationship on which the convection
model can be constructed.  The details of how this model can be constructed are given in \citet{augustson18}.  The
parameters needed to see how this model can be leveraged to give estimates of the rotational and diffusive influence are

\vspace{-0.25truein}
\begin{center}
  \begin{align}
     z^3 &= \frac{k^2}{k_z^2}, \quad q = \frac{N_0}{N}, \quad
    O = q \sqrt{\frac{3}{2}}\frac{\cos{\theta}}{5\pi\Roc}= q O_0, \nonumber\\
    K &= q \frac{\kappa k_z^2}{N_0} = q K_0,\quad
    V = q \frac{\nu k_z^2}{N_0} = q V_0, \label{eqn:equivalencies}
  \end{align}
\end{center}

\noindent where $k$ is the magnitude of the wavevector characterizing the mode that maximizes the heat flux, $k_z$ is
its vertical component, and $\theta$ is the colatitude. Note that the variation of the superadiabaticity for this system
is given by $\epsilon = H_P \beta/T$, meaning that $N^2 = g \alpha_T T \epsilon/H_P$, where $H_P$ is the pressure scale
height and $N_0$ is the buoyancy flux of the nondiffusive and nonrotating system provided by the Malkus-Howard
convection theory, $\beta$ is the potential temperature gradient, $\kappa$ is the thermal diffusion coefficient, $\nu$
is the viscous diffusion coefficient. The convective Rossby number is

\vspace{-0.25truein}
\begin{center}
  \begin{align}
    \Roc= \frac{\rmv_0}{2\Omega_0 \ell_0}=\frac{\sqrt{6}N_0}{10\pi\Omega_0},\label{eqn:rossby}
  \end{align}
\end{center}

\noindent where $\Omega_0$ is the constant angular velocity of the system and use has been made of the fact that the
characteristic velocity of the nonrotating system $\rmv_0$ is derived from the nonrotating and nondiffusive case as
$\rmv_0 = s_0/k_0 = \sqrt{6} N_0\ell_0/\left(5\pi\right)$, where $s_0$ is the growth rate derived from the dispersion
relationship in the nondiffusive and nonrotating case. Two relevant equations are the dispersion relationship linking
the normalized growth rate $\hat{s}=s/N_0$ to $q$ and $z$, and the heat flux $F$ to be maximized with respect to $z$

\vspace{-0.25truein}
\begin{center}
  \begin{align}
    &\left(\hat{s} \!+\! K_0 q z^3\right)\!\! \left(\! z^3\!\left(\hat{s}\!+\! V_0 q z^3\right)^2 \!\!+\! 4
      O_0^2q^2\!\right)\nonumber\\
    &-\! \left(z^3\!-\! 1\right)\!\!\left(\hat{s}\!+\! V_0 q z^3\right)\!=\!0, \label{eqn:fullcharq}\\
    &\frac{F}{F_0} = \frac{1}{q^3} \left[\frac{\hat{s}^3}{z^3}+V_0 q\hat{s}^2\right].\label{eqn:heatfluxq}
  \end{align}
\end{center}

To assess the scaling of the superadiabaticity, the velocity, and the horizontal
wavevector, a further assumption must be made in which the maximum heat flux is invariant to any parameters, namely that
$\max{\left[F\right]}=F_0$ so the heat flux is equal to the maximum value $F_0$ obtained in the Malkus-Howard turbulence
model for the nonrotating case.  Therefore, building this convection model consists of three steps: deriving a
dispersion relationship that links $\hat{s}$ to $q$ and $z$, maximizing the heat flux with respect to $z$, and assuming
an invariant maximum heat flux that then closes this three variable system.

\section{Convection Model}

In the case of planetary and stellar interiors, the viscous damping timescale is generally longer than the convective
overturning timescale (e.g., $V_0\ll N_0$).  Thus, the maximized heat flux invariance is much simpler to treat.
In particular, the flux invariance condition under this assumption is then

\vspace{-0.25truein}
\begin{center}
  \begin{align}
    \frac{\max{\left[F\right]}}{F_0} &=\left.\frac{\hat{s}^3}{q^3z^3}+\frac{V_0\hat{s}^2}{q^2}\right|_{\mathrm{max}}\approx
                                       \left.\frac{\hat{s}^3}{q^3z^3}\right|_{\mathrm{max}} =1 \nonumber\\
    &\implies \hat{s}=q z + \mathcal{O}(V_0/N_0).\label{eqn:maxndhf}
  \end{align}
\end{center}

One primary assumption of this convection model is that the magnitude of the velocity is defined as the ratio of the
maximizing growth rate and wavevector. With the above approximation, the velocity amplitude can be defined
generally. The velocity relative to the nondiffusive and nonrotating case scales as

\vspace{-0.25truein}
\begin{center}
  \begin{align}
    \frac{\rmv}{\rmv_0} &= \left(\frac{5}{2}\right)^{\frac{1}{6}}\frac{\hat{s}}{q z^{3/2}} = \left(\frac{5}{2}\right)^{\frac{1}{6}} z^{-\frac{1}{2}}.\label{eqn:vsteve}
  \end{align}
\end{center}

To find the scaling of the heat flux maximizing wavevector $k = z^{3/2}$ and the superadiabaticity
$\epsilon/\epsilon_0=q^{-2}$, one may find the implicit wavevector derivative of the growth rate $\hat{s}$ from Equation
\ref{eqn:fullcharq} and equate it to the derivative of the heat flux $\partial F/\partial z = \hat{s}/z$, which neglects
the heat flux arising from the viscous effects.  Using the heat-flux invariance, e.g. letting $\hat{s} = qz$, the
constraining dispersion relationship (Equation \ref{eqn:fullcharq}) can be manipulated to solve for $q$ as a function of
$z$.  Substituting this solution into the equation resulting from the flux maximization yields an equation solely for
the wavevector $z$:

\vspace{-0.25truein}
\begin{center}
  \begin{align}
    &z^3\! \left(V_0 z^2\!+\! 1\right)^2\times\nonumber\\
    &\left[3V_0 K_0 z^4\!\left(2 z^3\!-\! 3\right)+ z^2\!\left(V_0\!+\!
      K_0\right)\!\left(4z^3\!-\! 7\right)\!+\! 2 z^3\!-\! 5\right]\!-\nonumber\\
    &\frac{6 \cos^2{\!\theta}}{25 \pi^2 \Ro_{\mathrm{c}}^2}\!\! \left[K_0\left(3V_0z^5\!+\! z^3\!+\! 2\right)\!+\!
      V_0\left(5z^3\!-\! 2\right)\!+\! 3z\right]\!=\!0.\label{eqn:zeqndiff}
  \end{align}
\end{center}

\section{Wave Excitation and Energy Flux}\label{sec:waves}

\subsection{Gravito-Inertial Waves}

Internal gravito-inertial waves (GIWs) are a class of fluid waves where buoyancy and the Coriolis acceleration serve as
the restoring forces \citep[e.g.,][]{lee97,dintrans00,ballot10,prat16}. Within stars and planets with convectively
stable regions and potentially convective regions, each class of possible waves have frequency ranges over which they
can propagate, permitting nonlocal transport through radiative damping, corotation resonances, as well as through
nonlinear wave breaking \citep[e.g.,][]{schatzman93,mathis08,rogers15,pincon17}.  Considering local approximations to
these waves, the following can be seen. Pure gravity waves may propagate in stable regions if their frequency ($\omega$)
is less than the Brunt-V\"{a}is\"{a}l\"{a} frequency, and they are otherwise evanescent. Pure inertial waves may
propagate in rigidly-rotating convective regions when their frequency is less than twice the local rotation rate. In
convectively stable regions where rotation is important, GIWs may propagate if their frequency falls within the range
between $\omega_{-}$ and $\omega_{+}$ \citep[e.g.,][]{gerkema05},

\vspace{-0.25truein}
\begin{center}
  \begin{align}
	\omega_\pm \!=\! \frac{1}{\sqrt{2}}\sqrt{N_r^2\! +\! f^2\! +\! f_{ss}^2 \pm \!\sqrt{\left(N_r^2\! +\! f^2 \!+\! f_{ss}^2\right)^2 - \left(2N_r f\right)^2}}, \label{eqn:giwdisper}
  \end{align}
\end{center}

\noindent where $N_r$ is the Brunt-V\"{a}is\"{a}l\"{a} frequency in the stably stratified region,
$f=2\Omega_0\cos{\theta}$, $f_{ss}=2\Omega_0\sin{\theta}\sin{\psi}$, and $\psi$ is an angle in the plane transverse to
the local effective gravity vector.  At the pole in a convectively stable region, this implies that the frequency must
be between $2\Omega_0$ and $N_r$ for the wave to propagate, where the Brunt-V\"{a}is\"{a}l\"{a} frequency is typically
much larger than the rotational frequency in the radiative core of late-type stars and the radiative envelope of
early-type stars \citep[e.g.,][]{aerts10}. More generally, at other latitudes, the hierarchy of extremal propagative
wave frequencies satisfy the inequality $\omega_{-}<2\Omega_0<N_r<\omega_{+}$.  As these waves propagate, the
Brunt-V\"{a}is\"{a}l\"{a} frequency varies, for instance it becomes effectively zero in the convection zone. This
implies that waves in the frequency range $\omega<2\Omega_0$ are classified as sub-inertial GIWs in stable regions,
becoming pure inertial waves in convective regions. Waves in the frequency range $\omega\ge 2\Omega_0$ are classified as
super-inertial GIWs in the stable region become evanescent in the convective region as pure gravity waves.

\subsection{Wave Excitation Mechanisms}

One mechanism for exciting waves arises from stochastic motions in the convection zone due to Reynolds stresses
interacting with evanescent super-inertial GIWs or propagating inertial and sub-inertial waves as well as directly
exciting those waves when the convective flows penetrate into radiative zones. Such excitation mechanisms have been
captured in experiments and detected observationally \citep[e.g.,][]{neiner12a,aerts15,lebars15}. Moreover, the
convective excitation of waves and their associated transport processes have been examined extensively both through
theoretical modeling \citep[e.g.,][]{press81,belkacem09,pincon16} and through numerical simulations of convective flows
interacting with a stable layer \citep[e.g.,][]{rogers13,alvan15,lecoanet15}. Turbulence in convective regions is
modified by rotation and the rate of downscale energy transfer is reduced. In regimes of rapid rotation (e.g., low
convective Rossby number), strongly anisotropic flows develop that result from nonlinear interactions of propagative
inertial waves \citep[e.g.,][]{davidson13}. Such dynamics take place whether or not there is a stably stratified
region. However, when a stable zone is present, there are two distinct regimes: one where turbulence weakly influenced
by rotation is coupled with evanescent super-inertial GIWs and another where turbulence is strongly influenced by
rotation and is thus intrinsically and strongly coupled with sub-inertial GIWs
\citep[e.g.,][]{galtier03,clarkdileoni14,mathis14}. Nevertheless, these GIW and convection interactions and the
transport processes arising from them have not yet been taken into account. However, some steps toward incorporating
them into stellar models have already been achieved.  For instance, there has been a substantial effort to account for
the transport of angular momentum and chemical species by pure gravity waves in pre-main-sequence stars to stars in the
penultimate pre-supernovae state \citep[e.g.,][]{talon08,charbonnel13,fuller18}.

\subsection{Interfacial Gravito-Inertial Wave Flux Estimates}

There are many models for estimating the magnitude of the gravity wave energy flux arising from the waves excited by
convective flows. One of the first and most straightforward of such estimates is described in \citet{press81}, where the
wave flux across an interface connecting a convective region to a stable zone is computed by matching their respective
pressure perturbations at that interface. Because the wave excitation occurs at an interface, the pressure perturbations
are more important than the Reynolds stresses of the flows. What is more, the model assumes that the convective source
is a delta function in space and time. So, the model permits only a single horizontal spatial scale $2\pi/k_c$ and a
single time scale for the convection $2\pi/\omega_c$ that also selects the depth of the transitional interface where
$N(r)=\omega_c$ for gravity waves, where $\omega_c = \omega_0 /\sqrt{z}$ with $\omega_0 = 2\pi \rmv_0/\ell_0$, which
lends itself well to the above convection model. This approach yields a wave flux proportional to the product of the
convective kinetic energy flux and the ratio of the wave frequency to the Brunt-V\"{a}is\"{a}l\"{a} frequency in the
case of gravity waves.

The convective model established above captures some aspects of the influence of rotation on the convective
flows. Therefore, the impact of the Coriolis force on the stochastic excitation of GIWs can be evaluated. In this
context, recent work has established an estimate of the GIW flux \citep{andre17}.  It can be used to
estimate the rotational scaling of the amplitude of the wave flux arising from the modified properties of the convective
driving.  In \citet{andre17}, the vertical GIW energy flux is computed from the horizontal average of
the product of the vertical velocity and pressure perturbation that, given the linearization of the Boussinesq equations
for monochromatic waves propagating in a selected horizontal direction, can be evaluated to be

\vspace{-0.25truein}
\begin{center}
  \begin{align}
    F_z=\frac{1}{2}\rho_0 \frac{\omega^2-f^2}{\omega k_\perp^2} k_z \rmv_w^2,
  \end{align}
\end{center}

\noindent where $\rmv_w$ is the magnitude of the vertical velocity of the wave.

Following \citet{press81}, further assumptions are necessary to complete the estimate of the wave flux.  The convection
is turbulent. So the fluctuating part of the velocity field is of the same order of magnitude as the convective eddy
turnover velocity $\rmv_c \approx \omega_c/k_c$, which implies that convective pressure perturbations are approximately
$P_c = \rho_0 \rmv_c^2$. Assuming that the pressure is continuous across the interface between the convectively stable
and unstable regions, the horizontally-averaged pressure perturbations of the propagating waves excited at the interface
must then be equal to the turbulent pressure on the convective side of the interface. Those pressure perturbations
follow from the solution for the vertical velocity and the nondiffusive Boussinesq equations. Flows in a gravitationally
stratified convective medium tend to have an extent in the direction of gravity that is much larger than their extent in
the transverse directions.  Therefore, the horizontal wavenumber of the convective flows is much greater than the
vertical wavenumber. Within the context of the single-mode convection model derived above, this implies that
$k_{\perp,c} \approx \omega_c/\rmv_c$. For efficient wave excitation, the frequency of the wave needs to be close to the
source frequency \citep{press81,lecoanet13}, which means that the horizontal scale of the waves will be similar to that
of the convection. Therefore, the spectral density in frequency space of the wave energy flux is approximately

\vspace{-0.25truein}
\begin{center}
  \begin{align}
    F_z\approx\frac{1}{2}\frac{\rho_0 \rmv_c^3 \omega^2\left[\omega^2 f_{ss}^2 + \left(N_r^2-\omega^2\right)\left(\omega^2-f^2\right)\right]^{\frac{1}{2}}}{\omega^2 f_s^2 + \left(N_r^2-\omega^2\right)\left(\omega^2-f^2\right)}.
  \end{align}
\end{center}

Finally, taking the ratio of the rotation-influenced flux to the flux in the nonrotating case to better isolate the
changes induced by rotation, assuming that the Brunt-V\"{a}is\"{a}l\"{a} frequency is not directly impacted by rotation,
and at a fixed wave frequency $\omega$, one has that

\vspace{-0.25truein}
\begin{center}
  \begin{align}
    \frac{F_z}{F_0} \approx \left(\frac{\rmv}{\rmv_0}\right)^{\!\!3} \!\frac{\omega\!\left\{\!\left(N_r^2\!-\!\omega^2\right)\!\left[\omega^2
    f_{ss}^2 \!+\! \left(N_r^2\!-\!\omega^2\right)\left(\omega^2\!-\! f^2\right)\!\right]\!\right\}^{\!\frac{1}{2}}}{\omega^2
    f_s^2 \!+\! \left(N_r^2\!-\! \omega^2\right)\left(\omega^2\!-\! f^2\right)},
  \end{align}
\end{center}

\noindent with $f_{s} = 2\Omega_0\sin{\theta}$ and $f_{ss}$ defined as above. To make this a bit more parametrically
tractable, one can normalize the wave frequency as $\sigma=\omega / N_r$, and cast the rotational terms into a product
of the stiffness of the transition $S=N_r/N_0$, with the convective Rossby number as defined above in \S
\ref{sec:genframe} with Equations \ref{eqn:equivalencies} and \ref{eqn:rossby}.  Doing so yields

\vspace{-0.25truein}
\begin{center}
  \begin{align}
    &\frac{F_z}{F_0}\! \approx\! \left(\frac{\rmv}{\rmv_0}\right)^{\!\!3} \!\left[\Ro_{\mathrm{w}}^{-2}\sin^2{\!\theta}\!+\! \left(\sigma^{-2}\!-\!1\right)\!\!\left(\!1\!-\!\Ro_{\mathrm{w}}^{-2}\cos^2{\!\theta}\right)\!\right]^{\!-1}\nonumber\\
    &\left[\!\left(\sigma^{-2}\!-\!1\right)\!\Ro_{\mathrm{w}}^{-2}\sin^2{\!\theta}\sin^2{\!\psi}\!+\!\left(\sigma^{-2}\!-\!1\right)^2\!\left(1\!-\!\Ro_{\mathrm{w}}^{-2}\cos^2{\!\theta}\right)\!\right]^{\!\frac{1}{2}}\!\!\!,
  \end{align}
\end{center}

\begin{figure}[t!]
  \begin{center}
    \vspace*{1cm}\includegraphics[width=0.45\textwidth]{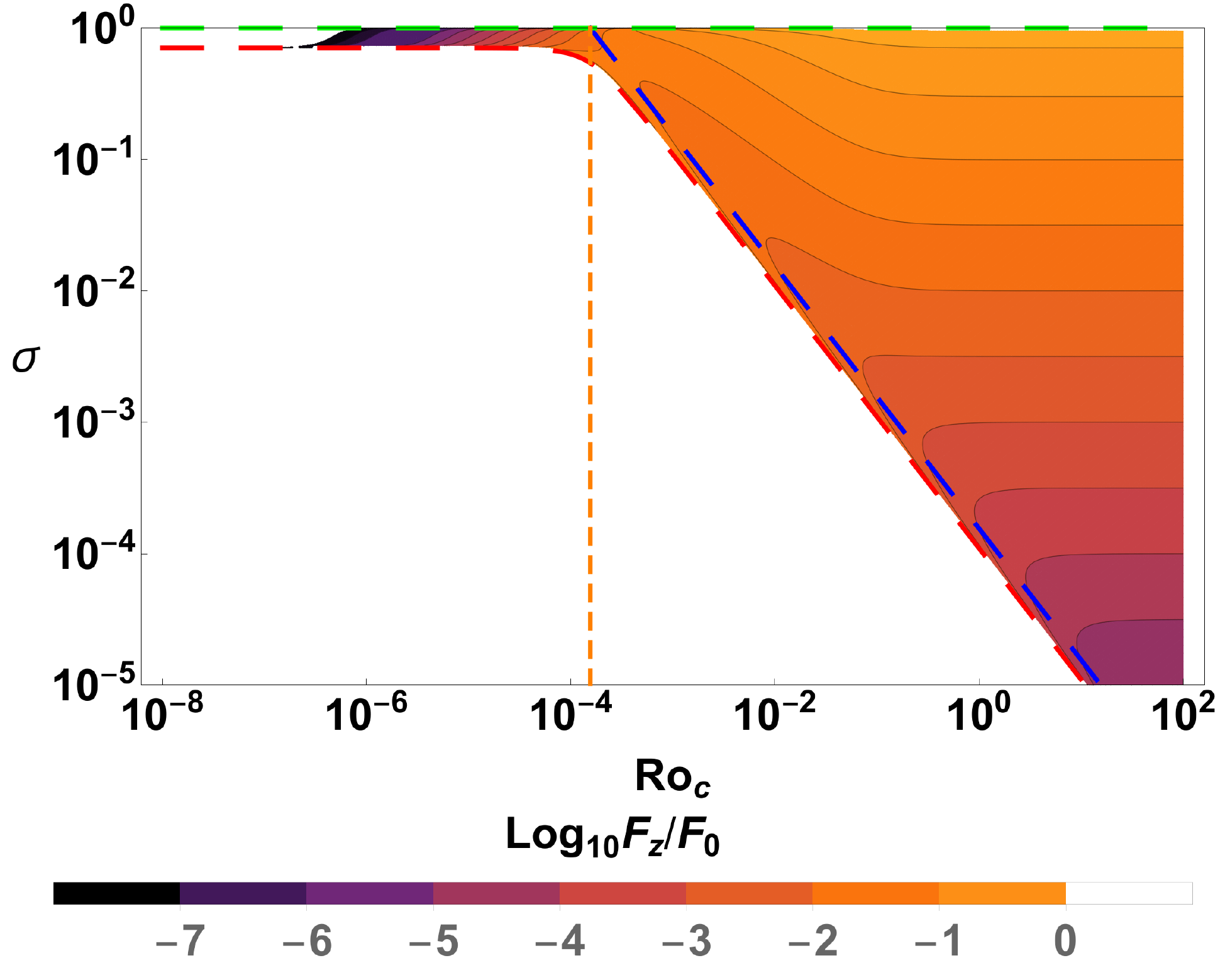} 
    \caption{Convective Rossby number dependence of the ratio of the interfacial gravity wave flux excited by rotating
      convection relative to the nonrotating case $F_z/F_0$ for the nondiffusive convection model at $\theta=\pi/4$,
      with an interface stiffness of $S = 10^{3}$ and a horizontal direction of $\psi=\pi/2$. The red dashed line
      indicates the lower frequency cutoff $\sigma_{-}$, the green dashed line indicates the upper cutoff frequency of
      $\sigma=1$ since the wave flux is being compared to a non-rotating case, whereas the blue dashed line indicates a
      wave Rossby number of $\Rowt=1$.  The vertical dashed orange line indicates the critical convective Rossby
      number. }\label{fig:wave_scaling}
  \end{center}
\end{figure}

\noindent where $\Row=\omega/2\Omega_0=5\pi\sigma \Ro_{\mathrm{c}} S/\sqrt{6}$ and the wave Rossby number is
$\Rowt=\Row/\sqrt{\sin^2{\theta}\sin^2{\psi}+\cos^2{\theta}}$.

This is depicted in Figure \ref{fig:wave_scaling}, where the colored region exhibits the magnitude of the logarithm of
the flux ratio.  An interfacial stiffness of $S=10^3$ is chosen as it is a rough estimate of the potential stiffness in
most stars, being the ratio of the buoyancy time-scale in the stable region to the convective overturning time. The
choice of latitude determines the width of the frequency band of sub-inertial waves, where it is a minimum at the pole
and maximum near the equator, with $\theta=\pi/4$ here and $\theta\approx\pi/2$ shown in \citet{augustson18}. The
direction of $\psi=\{-\pi/2,\pi/2\}$ is chosen as it represents the maximum value of the flux ratio for the choice of
other parameters and represents the waves travelling toward either of the poles as the flux ratio is an even parity
function of the horizontal direction. Specifically, the poleward wave flux ratio is greater than the other extremal
choice of the prograde or retrograde wave flux ratios. In particular, given the range of $\omega_{\pm}$, there are no
sub-inertial waves in the prograde or retrograde propagation case ($\psi=\{0,\pi\}$, respectively), whereas the
super-inertial waves may still propagate with roughly the same frequency range.  The white region corresponds to the
domain of evanescent waves for a given Rossby number with frequencies below the lower cut-off frequency ($\sigma_{-}$,
dashed red line) for propagating GIWs. At frequencies above this threshold there is a frequency dependence of the flux
ratio until reaching the upper cut-off where $\sigma=\omega/N_r = 1$, which arises due to the domain of validity when
comparing GIW to gravity wave fluxes. Indeed, gravity waves may propagate if $\omega<N_r$, whereas super-inertial GIWs
may propagate even when $N_r<\omega<\omega_{+}$. The transition between super-inertial and sub-inertial waves is
demarked with the dashed green line, with super-inertial waves for $\Rowt>1$ and sub-inertial waves for $\Rowt<1$. Here,
interfacially-excited super-inertial waves exhibit both a frequency and convective Rossby number
dependence. Specifically, the wave flux decreases algebraically with frequency at a fixed convective Rossby number and
have a reduced flux for convective Rossby numbers below unity.  The interfacially-excited sub-inertial waves possess a
small frequency domain at a fixed convective Rossby number over which they are propagative.  The sub-inertial wave flux
increases with decreasing convective Rossby number until a critical convective Rossby number
$\Ro_{c,\mathrm{crit}} = \sqrt{6}/\left(5\pi S\right)$ as depicted by the vertical dashed orange line in Figure
\ref{fig:wave_scaling}. Below this critical convective Rossby number, the sub-inertial wave flux decreases and their
frequency domain is further restricted until it vanishes entirely and there are no propagative super-inertial waves .
The effect of the stiffness is to lower (raise) the value of the critical convective Rossby number for larger (smaller)
values of $S$, which corresponds to the ratio of the buoyancy time-scale in the radiative zone to the convective
overturning time. This may have important consequences for the wave-induced transport of angular momentum during the
evolution of rotating stars.  In particular, the convective Rossby number can vary by several orders of magnitude over a
star's evolution from the PMS to its ultimate demise. Moreover, it can vary internally as a function of radius due to
the local amplitude of the convective velocity and due to transport processes, angular momentum loss through winds, and
structural changes that modify the local rotation rate \citep[e.g.,][]{landin10,mathis16,charbonnel17}.

\subsection{Reynolds Stress Contributions to GIW Amplitudes}

So far, the wave excitation mechanism has been considered to be interfacial, as in \citet{press81}.  Yet this estimate
omits wave flux linked to their excitation by convective Reynolds stresses \citep[e.g.,
][]{belkacem09,samadi10,lecoanet15}. The amplitude and the wave flux of the GIWs may be estimated using the results of
\citet{mathis14}, where the impact of rotation on the waves is treated coherently.  In a means similar to
\citet{goldreich90} and \citet{lecoanet13}, although with a greater degree of computational complexity, one may derive
the wave amplitudes for GIWs in a f-plane. One must first find solutions to the homogeneous Poincar\'{e} equation for
the GIWs and then use linear combinations of those solutions to construct solutions to the forced Poincar\'{e} equation
as in Appendix B of \citet{gerkema05} and Equation 27 of \citet{mathis14}. Those solutions can then be employed to
construct asymptotic expressions for the wave amplitude for a given forcing function in the convection zone as in
Equations 28 and 29 of \citet{mathis14}. This can be evaluated analytically using the properties of the Fourier
transform, the convolution theorem, and the properties of Dirac delta functions (see \citet{augustson18} for details).

Precisely defining the spectral properties of the Reynolds stresses in rotating thermal convection is a difficult
task. The symmetries of homogeneous and isotropic stirred turbulence do not generally exist in these systems. Thus, for
simplicity and as a first approximation, the turbulence model presented in \citet{lecoanet13} is adopted here. In this
model, the convection is treated as a Kolmogorov-Heisenberg spectrum of eddies, with the eddies spanning the size
spectrum from a maximum cut off size of $H$ to eddies of size $h<H$ and with a large-scale turnover frequency of
$\omega_c$ and eddy time-scale with $\omega_e>\omega_c$.  The eddy velocity thus scales as
$u_h = u_c (h/H)^{1/3}=u_c(\omega_e/\omega_c)^{-1/2}$, where $\omega_e = u_h/h$ and $u_c$ is the rms convective
velocity.  Thus, the Reynolds stresses due to a single eddy are considered to be
$\widehat{\rmv_i\rmv_j} \propto h^3 u_h^2$. Given this prescription of convection, one may estimate the magnitude of the
forcing function due to the convective Reynolds stresses in the forced Poincar\'{e} equation that describes the wave
motion.  Specifically, the amplitude of the waves may be estimated, which leads to subsequent scaling relationships for
the spectral density $\partial^2 P/\partial \ln{\omega}\partial\ln{k_\perp}$ of the wave flux that has been integrated
in horizontal wavevector $k_\perp$. These scaling relationships for the gravity wave flux have shown to hold up well in
numerical simulations \citep[e.g.,][]{couston18}. With a turbulence model in hand, one may closely follow
\citet{lecoanet13} with some new assumptions about the mode number density to find the scaling for the wave flux ratio,
where it is left to the reader to seek out the details of the derivation in \citet{augustson18}. Note that again
$S=N_r/N_0$ and $\omega_c^2 = 6\rmv^2N_0^2/(25\pi^2\rmv_0^2)$. Thus, following the approach in the interfacial
excitation case for nondimensionalizing the Reynolds stress driven flux ratio estimate, one has that

\begin{center}
  \begin{align}
    &\frac{F_{z,j}}{F_0} =\frac{n_z\int \!\!dk_\perp\mathrm{d}_j G_j^2}{n_0p^2\!\left.\int \!\!dk_\perp\mathrm{d}_j G_j^2\right|_{0}}\left(\frac{\rmv}{\rmv_0}\right)^{\frac{9}{2}}\left(1-\Ro_{\mathrm{w}}^{-2}\cos^2{\theta}\right)^3\nonumber\\
    &\times\left(\left[\ell_sL^{-1}-6\ell_0L^{-1}/\left(25\pi^2S^2\right)\right]\sigma^{-2}-1\right)\nonumber\\
    &\times\left\{\left[\ell_sL^{-1}-6\ell_0L^{-1}/\left(25\pi^2S^2\right)\right]\left(1-\Ro_{\mathrm{w}}^{-2}\cos^2{\theta}\right)
      \sigma^{-2}\right.\nonumber\\
    &\qquad+\widetilde{\Ro}_{\mathrm{w}}^{-2}-1\big\}^{-1},
  \end{align}
\end{center}

\noindent where $\Ro_\mathrm{w} = \omega/2\Omega_0=\sqrt{25/6}\pi \Ro_{\mathrm{c}}S \sigma$ and $j=\rmsub$ or $\rmsup$
for sub-inertial and super-inertial waves respectively, and
$p=\sqrt{1-\widetilde{\Ro}_{\mathrm{w}}^{-2}}/\left(1-\Ro_{\mathrm{w}}^{-2}\cos^2{\theta}\right)$. The ratio of the mode
number densities is

\begin{center}
  \begin{align}
&\frac{n_z}{n_0}=\left[\ell_sL^{-1}\left|\frac{\sigma^{-2}-1}{1-\Ro_{\mathrm{w}}^{-2}\cos^2{\theta}}+\frac{\delta^2\Ro_{\mathrm{w}}^{-2}}{\cos^2{\theta}}\right|^{\frac{1}{2}}\right.\nonumber\\
&\qquad\left.+\ell_0L^{-1}\left|\frac{\delta^2\Ro_{\mathrm{w}}^{-2}}{\cos^2{\theta}}-\frac{6\rmv^2\sigma^{-2}/\left(25\pi^2\rmv_0^2S^{2}\right)+1}{1-\Ro_{\mathrm{w}}^{-2}\cos^2{\theta}}\right|^{\frac{1}{2}}\right]\nonumber\\
    &\times\!\left[\ell_sL^{-1}\left(\sigma^{-2}-1\right)^{\frac{1}{2}}+\ell_0L^{-1}\left(6\sigma^{-2}/\left(25\pi S^2\right)+1\right)^{\frac{1}{2}}\right]^{-1}\!\!\!\!,
  \end{align}
\end{center}

\noindent where the horizontal phase shift is
$\delta=\sin{\theta}\cos{\theta}\sin{\psi}/\left(1-\Ro_{\mathrm{w}}^{-2}\cos^2{\theta}\right)$. The Fourier transforms
and convolutional integrals of the asymptotic eigenfunctions for the GIWs and the convective source function result in
an integral of the wave-turbulence interaction structure functions $n_zd_j G_j$ with respect to the horizontal wave
number. These functions are composed of a mode excitation depth $d_j$, a wave structure function $G_j$, and a mode
number density $n_z$. For the super-inertial waves, this is

\begin{center}
  \begin{align}
    &Q_\rmsup=\int_{2\pi L^{-1}}^{k_\perp} d{k'}_\perp d_\rmsup G_\rmsup^2\nonumber\\
    &=\left[p^2\!\left(2\delta\!+\!\zeta\right)^2\!+\!\left(1\!+\!p^2\!-\!\delta^2\!-\!\delta\zeta\right)\!\right]\times\nonumber\\
    &\qquad\int_{2\pi L^{-1}}^{k_\perp}d{k'}_\perp \ell_0e^{-{k'}_\perp\left( p \ell_0+2\Delta\right)}\nonumber\\
    &=\left[p^2\!\left(2\delta\!+\!\zeta\right)^2\!+\!\left(1\!+\!p^2\!-\!\delta^2\!-\!\delta\zeta\right)\!\right]\left[\ell_s\ell_0^{-1} \Delta+p\right]^{-1}\!\!\times\nonumber\\
    &\qquad\left[e^{-2\pi L^{-1}\left(\ell_s\Delta+p\ell_0\right)}-e^{-k_\perp \left(\ell_s\Delta+p\ell_0\right)}\right].
  \end{align}
\end{center}

\noindent For the sub-inertial waves, this is

\begin{center}
  \begin{align}
    &Q_\rmsub=\int_{2\pi L^{-1}}^{k_\perp} d{k'}_\perp d_\rmsub G_\rmsub^2=\int_{2\pi L^{-1}}^{k_\perp} d{k'}_\perp \ell_0\times\nonumber\\
    &\left[p^2\!\left(2\delta\!+\!\zeta\right)^2\!\cos^2{\!\left(k_\perp\ell_s\Delta\right)}\!+\! \left(\delta^2\!+\!\delta\zeta\!+\!p^2\!-\!1\right)^2\!\sin^2{\!\left(k_\perp\ell_s\Delta\right)}\right]\nonumber\\
    &=\frac{\left(k_\perp L\!-\!2\pi\right)}{2\ell_0^{-1}L}\!\left[p^2\!\left(2\delta\!+\!\zeta\right)^2\!+\!\left(\delta^2\!+\!\delta\zeta\!+\!p^2\!-\!1\right)^2\right]+\nonumber\\
    &\qquad\left[p^2\!\left(2\delta\!+\!\zeta\right)^2\!-\!\left(\delta^2\!+\!\delta\zeta\!+\!p^2\!-\!1\right)^2\right]\times\nonumber\\
    &\qquad\qquad\frac{\sin\left[2k_\perp \ell_s\Delta\right]\!-\!\sin\left[4\pi\ell_sL^{-1}\Delta\right]}{4\ell_s\ell_0^{-1}\Delta},
  \end{align}
\end{center}

\noindent where $\ell_0=z_2-z_c$, $\ell_s=z_c-z_1$, $L=z_2-z_1$, are the depths of the convection zone, stable layer,
and whole domain respectively, with $z_c$ being the point of sharp variation in the Brunt-V\"{a}is\"{a}l\"{a}
frequency modelled here using an Heaviside distribution. Here $\zeta$ is an anisotropy parameter for the convective
Reynolds stresses, with $\zeta=1$ for isotropic flows, and

\vspace{-0.25truein}
\begin{center}
  \begin{align}
    \Delta= \ell_s\left|\frac{\sigma^{-2}-1}{1-\Ro_{\mathrm{w}}^{-2}\cos^2{\theta}} +\frac{\delta^2\Ro_{\mathrm{w}}^{-2}}{\cos^2{\theta}}\right|^{\frac{1}{2}},
  \end{align}
\end{center}

\noindent is the effective wave cavity size for a given frequency.

\begin{figure*}[ht!]
  \begin{center}
    \includegraphics[width=\textwidth]{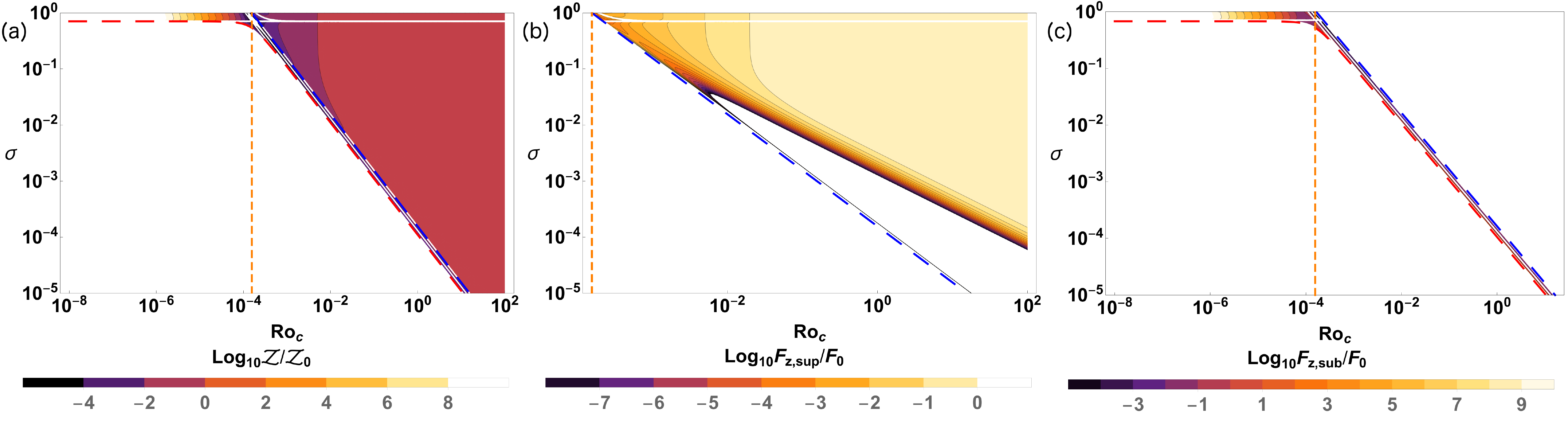} 
    \caption{Convective Rossby number dependence of the ratio of the GIW flux excited by rotating convection relative to
      the nonrotating case $F_z/F_0$ for the nondiffusive convection model at $\theta=\pi/4$, with an interface
      stiffness of $S = 10^{3}$, a horizontal direction of $\psi=\pi/2$, in a symmetric domain where
      $\ell_s=\ell_0=L/2$, and with $\zeta=1$. (a) The flux $\mathcal{Z}/\mathcal{Z}_0$ is shown without the influence
      of the $Q_j$, which serves to highlight its inclusion in the next two panels. In all panels, the dashed red line
      denotes the lower wave frequency cutoff $\sigma_{-}$, whereas the dashed blue line corresponds to $\Rowt =1$. The
      vertical dashed orange line indicates the critical convective Rossby number. The upper cutoff is $\sigma=1$ as
      this is the maximum frequency for which the rotating and non-rotating cases can be compared. (b) The total wave
      flux $F_{z,\rmsup}/F_0$ for the super-inertial waves, illustrating the truncation of the frequency spectrum due to
      $Q_\rmsup$. (c) The total wave flux $F_{z,\rmsup}/F_0$ for the sub-inertial waves, exhibiting the increasing flux
      enhancement when $\sigma$ is such that $\Rowt\approx 1$ and $\Ro_{\mathrm{c}}<Ro_{\mathrm{c,crit}}$.}\label{fig:wave_scaling2}
  \end{center}
\end{figure*}

Figure \ref{fig:wave_scaling2} separately shows the flux ratio of convectively-excited GIWs relative to gravity waves in
a nonrotating limit and omitting the factor of $Q_j/Q_j|_{0}$, denoted as $\mathcal{Z}/\mathcal{Z}_0$, as well as the
flux ratios that include this prefactor for the super-inertial and sub-inertial waves. The parameters are chosen to be
the same as for the interfacial case (Figure \ref{fig:wave_scaling}), with the additional choice of a symmetric domain
and assuming isotropic turbulence so that $\zeta=1$. Note that the effect of $\zeta>1$ is to lower the GIW flux relative
to the nonrotating case. In Figure \ref{fig:wave_scaling2}(a), two distinct regions appear to the left and to the
right of the $\Rowt=1$ line, which is drawn in blue.  Waves to the right of this line have $\Rowt>1$ and thus correspond
to super-inertial waves, and those on the left correspond to sub-inertial waves. The dashed red line is the lower cutoff
frequency for propagative GIWs.  Again there is the critical convective Rossby number, depicted by the vertical dashed
orange line, below which there are no super-inertial waves and only a small frequency band of sub-inertial waves are
permitted. From Figures \ref{fig:wave_scaling2}(a) and (b), it is clear that the flux ratio decreases for super-inertial
waves as the convective Rossby number decreases, just as in the interfacial excitation model discussed above and shown
in Figure \ref{fig:wave_scaling}. Moreover, this also indicates that the frequency range of the GIWs remains restricted
at lower Rossby numbers when comparing to the interfacially-excited GIWs.  Indeed, the convectively-excited
super-inertial waves have a more restricted frequency range due to the exponential behavior of the $Q_\rmsup$ factor
when compared to the interfacially-excited waves.  Particularly, the flux ratio drops below the lower cutoff value of
$10^{-8}$ and becomes vanishingly small in the white region between the dark lower bound and the dashed blue line.  The
convectively-excited sub-inertial GIW flux ratio is shown in Figure \ref{fig:wave_scaling2}(c), where it is evident that
the waves remain highly restricted to a narrow frequency band as is the case for the interfacially-excited sub-inertial
waves.  However, when the convective Rossby number is below the critical convective Rossby number, the
convectively-excited sub-inertial waves have an exponentially increasing flux ratio with decreasing Rossby number.  For
sub-inertial waves propagating in regions where the convective Rossby number is greater than that critical value, there
is a very narrow band of frequencies close to $\sigma=\sigma_{-}$ for which the flux ratio is very large in an apparent
resonance. Given the form of $Q_\rmsub$, this is expected due to the inverse dependence of $Q_\rmsub$ on $\Delta$.  The
super-inertial waves, on the other hand, have $Q_\rmsup\propto \left(\Delta + p\right)^{-1}$ and so do not have the same
peaked feature.  When further comparing the interfacially-excited and the convectively-excited GIWs, it appears that
there is a transition in frequency below which the convectively-excited super-inertial waves will have a greater flux
ratio than the interfacially excited ones. The sub-inertial waves, on the other hand, have a much richer suite of
behavior, with there being a transition in convective Rossby number where a similar exchange between the dominant
excitation mechanism occurs.  There is also a broader range of frequencies where the interfacially-excited sub-inertial
waves have an appreciable flux ratio, and yet a very strong resonant peak for the convectively-excited waves that dwarfs
the magnitude of the flux ratio of interfacially-excited waves. Such resonances are not without precedence, as they
appear in numerical simulations \citep{rogers13} and may be at the origin of stochastically excited sub-inertial waves
observed for some rapidly rotating stars \citet{neiner12a}.

\section{Summary and Discussion}\label{sec:final}

A simple model of rotating convection originating with \citet{stevenson79} has been extended to include thermal and
viscous diffusion for any convective Rossby number.  Moreover, a systematic means of developing such models for an
arbitrary dispersion relation have also been shown.  An explicit expression is given for the scaling of the horizontal
wavenumber in terms of the Rossby number and diffusion coefficients under the constraint that the values of the
diffusive time scale are less than the convective time scale (Equation \ref{eqn:zeqndiff}).  The scaling of the velocity
in terms of that wavenumber is also given.  As shown in \citet{augustson18}, these match the expressions given in
\citet{stevenson79}, as well as the numerical results found in the 3D simulations of \citet{kapyla05} and
\citet{barker14} asymptotically at low Rossby number and without diffusion.

As a first step in examining the impact of rotation on the stochastic excitation of GIWs, the convection model is
applied to the interfacial wave excitation paradigm developed in \citet{press81}.  Following the wave dynamics given in
\citet{mathis14} and \citet{andre17}, one can derive that the \citet{press81} model implies that the wave flux for
propagative waves should decrease asymptotically as $\Ro_{\mathrm{c}}^{3/5}$.  Also, for a given stiffness of the
transition between the radiative zone and the convection zone, the domain of allowed frequencies for propagative waves
steadily decreases as seen in Figure \ref{fig:wave_scaling}.  To better assess the influence of the convective Reynolds
stresses and of rotation on the GIWs, a third wave flux estimate is constructed using an explicit computation of the
amplitude for both the super-inertial and sub-inertial waves derived in \citet{mathis14}.  The convection model of
\citet{lecoanet13} is then invoked as means of estimating the Reynolds stresses and their variation with rotation.  It
is found again that the wave flux decreases with Rossby number (Figure \ref{fig:wave_scaling2}) due to the influence of
the convective velocities.  These fluxes also depend upon the horizontal wave number of the wave.  More strikingly,
however, the super-inertial waves additionally have a truncated frequency spectrum of waves capable of transporting
energy whereas the sub-inertial waves have a small range of frequencies centered around $\omega_0\propto N_rRo_c^{-1}$
over which their flux is greatly increased to levels similar to the nonrotating cases. In stark contrast to the
interfacially excited waves, the convectively-excited sub-inertial waves also exhibit an exponentially increasing wave
flux relative to the nonrotating case below a critical convective Rossby number. This may have substantial consequences
for the transport and mixing of angular momentum, chemical species, and heat in rotating stellar and planetary
interiors.

\section*{Acknowledgments} { K.~C. Augustson and
  S. Mathis acknowledge support from the ERC SPIRE 647383 grant and PLATO CNES grant at CEA/DAp-AIM. }

\bibliography{augustson_convection}

\begin{thebibliography}{62}
\providecommand{\natexlab}[1]{#1}

\bibitem[\protect\astroncite{{Aerts} \emph{et~al.}}{2010}]{aerts10}
{Aerts}, C., {Christensen-Dalsgaard}, J., \& {Kurtz}, D.~W. 2010,
  \emph{{Asteroseismology}}.

\bibitem[\protect\astroncite{{Aerts} \& {Rogers}}{2015}]{aerts15}
{Aerts}, C. \& {Rogers}, T.~M. 2015, \apjl, 806, L33.

\bibitem[\protect\astroncite{{Aerts} \emph{et~al.}}{2017}]{aerts17}
{Aerts}, C., {Van Reeth}, T., \& {Tkachenko}, A. 2017, \apjl, 847, L7.

\bibitem[\protect\astroncite{{Alvan} \emph{et~al.}}{2013}]{alvan13}
{Alvan}, L., {Mathis}, S., \& {Decressin}, T. 2013, \aap, 553, A86.

\bibitem[\protect\astroncite{{Alvan} \emph{et~al.}}{2015}]{alvan15}
{Alvan}, L., {Strugarek}, A., {Brun}, A.~S., {Mathis}, S., \& {Garcia}, R.~A.
  2015, \aap, 581, A112.

\bibitem[\protect\astroncite{{Andr{\'e}} \emph{et~al.}}{2017}]{andre17}
{Andr{\'e}}, Q., {Barker}, A.~J., \& {Mathis}, S. 2017, \aap, 605, A117.

\bibitem[\protect\astroncite{{Augustson} \& {Mathis}}{2018}]{augustson18}
{Augustson}, K.~C. \& {Mathis}, S. 2018, \apj, 869, 33.

\bibitem[\protect\astroncite{{Ballot} \emph{et~al.}}{2010}]{ballot10}
{Ballot}, J., {Ligni{\`e}res}, F., {Reese}, D.~R., \& {Rieutord}, M. 2010,
  \aap, 518, A30.

\bibitem[\protect\astroncite{{Barker} \emph{et~al.}}{2014}]{barker14}
{Barker}, A.~J., {Dempsey}, A.~M., \& {Lithwick}, Y. 2014, \apj, 791, 13.

\bibitem[\protect\astroncite{{Belkacem} \emph{et~al.}}{2009}]{belkacem09}
{Belkacem}, K., {Mathis}, S., {Goupil}, M.~J., \& {Samadi}, R. 2009, \aap, 508,
  345.

\bibitem[\protect\astroncite{{Brun} \& {Toomre}}{2002}]{brun02}
{Brun}, A.~S. \& {Toomre}, J. 2002, \apj, 570, 865.

\bibitem[\protect\astroncite{{Chandrasekhar}}{1961}]{chandrasekhar61}
{Chandrasekhar}, S. 1961, \emph{{Hydrodynamic and hydromagnetic stability}}.

\bibitem[\protect\astroncite{{Charbonnel} \emph{et~al.}}{2013}]{charbonnel13}
{Charbonnel}, C., {Decressin}, T., {Amard}, L., {Palacios}, A., \& {Talon}, S.
  2013, \aap, 554, A40.

\bibitem[\protect\astroncite{{Charbonnel} \emph{et~al.}}{2017}]{charbonnel17}
{Charbonnel}, C., {Decressin}, T., {Lagarde}, N., {Gallet}, F., {Palacios}, A.,
  \emph{et~al.} 2017, \aap, 605, A102.

\bibitem[\protect\astroncite{{Clark di Leoni}
  \emph{et~al.}}{2014}]{clarkdileoni14}
{Clark di Leoni}, P., {Cobelli}, P.~J., {Mininni}, P.~D., {Dmitruk}, P., \&
  {Matthaeus}, W.~H. 2014, Physics of Fluids, 26, 035106.

\bibitem[\protect\astroncite{{Couston} \emph{et~al.}}{2018}]{couston18}
{Couston}, L.-A., {Lecoanet}, D., {Favier}, B., \& {Le Bars}, M. 2018, Journal
  of Fluid Mechanics, 854, R3.

\bibitem[\protect\astroncite{Davidson}{2013}]{davidson13}
Davidson, P. 2013, \emph{Turbulence in Rotating, Stratified and Electrically
  Conducting Fluids} (Cambridge University Press).
\newblock ISBN 9781107434349.

\bibitem[\protect\astroncite{{Dintrans} \& {Rieutord}}{2000}]{dintrans00}
{Dintrans}, B. \& {Rieutord}, M. 2000, \aap, 354, 86.

\bibitem[\protect\astroncite{{Dintrans} \emph{et~al.}}{1999}]{dintrans99}
{Dintrans}, B., {Rieutord}, M., \& {Valdettaro}, L. 1999, Journal of Fluid
  Mechanics, 398, 271.

\bibitem[\protect\astroncite{{Fuller} \emph{et~al.}}{2015}]{fuller15b}
{Fuller}, J., {Cantiello}, M., {Stello}, D., {Garcia}, R.~A., \& {Bildsten}, L.
  2015, Science, 350, 423.

\bibitem[\protect\astroncite{{Fuller} \& {Ro}}{2018}]{fuller18}
{Fuller}, J. \& {Ro}, S. 2018, \mnras, 476, 1853.

\bibitem[\protect\astroncite{{Galtier}}{2003}]{galtier03}
{Galtier}, S. 2003, \pre, 68, 015301.

\bibitem[\protect\astroncite{{Garaud}}{2018}]{garaud18}
{Garaud}, P. 2018, Annual Review of Fluid Mechanics, 50, 275.

\bibitem[\protect\astroncite{{Garc{\'{\i}}a} \emph{et~al.}}{2007}]{garcia07}
{Garc{\'{\i}}a}, R.~A., {Turck-Chi{\`e}ze}, S., {Jim{\'e}nez-Reyes}, S.~J.,
  {Ballot}, J., {Pall{\'e}}, P.~L., \emph{et~al.} 2007, Science, 316, 1591.

\bibitem[\protect\astroncite{{Gerkema} \& {Shrira}}{2005}]{gerkema05}
{Gerkema}, T. \& {Shrira}, V.~I. 2005, Journal of Fluid Mechanics, 529, 195.

\bibitem[\protect\astroncite{{Glatzmaier} \& {Gilman}}{1982}]{glatzmaier82}
{Glatzmaier}, G.~A. \& {Gilman}, P.~A. 1982, \apj, 256, 316.

\bibitem[\protect\astroncite{{Goldreich} \& {Kumar}}{1990}]{goldreich90}
{Goldreich}, P. \& {Kumar}, P. 1990, \apj, 363, 694.

\bibitem[\protect\astroncite{{Howard}}{1963}]{howard63}
{Howard}, L.~N. 1963, Journal of Fluid Mechanics, 17, 405.

\bibitem[\protect\astroncite{{Jouve} \& {Ogilvie}}{2014}]{jouve14}
{Jouve}, L. \& {Ogilvie}, G.~I. 2014, Journal of Fluid Mechanics, 745, 223.

\bibitem[\protect\astroncite{{K{\"a}pyl{\"a}} \emph{et~al.}}{2005}]{kapyla05}
{K{\"a}pyl{\"a}}, P.~J., {Korpi}, M.~J., {Stix}, M., \& {Tuominen}, I. 2005,
  \aap, 438, 403.

\bibitem[\protect\astroncite{{Kichatinov}}{1986}]{kichatinov86}
{Kichatinov}, L.~L. 1986, Geophysical and Astrophysical Fluid Dynamics, 35, 93.

\bibitem[\protect\astroncite{{Kumar} \emph{et~al.}}{1999}]{kumar99}
{Kumar}, P., {Talon}, S., \& {Zahn}, J.-P. 1999, \apj, 520, 859.

\bibitem[\protect\astroncite{{Kupka} \& {Muthsam}}{2017}]{kupka17}
{Kupka}, F. \& {Muthsam}, H.~J. 2017, Living Reviews in Computational
  Astrophysics, 3, 1.

\bibitem[\protect\astroncite{{Landin} \emph{et~al.}}{2010}]{landin10}
{Landin}, N.~R., {Mendes}, L.~T.~S., \& {Vaz}, L.~P.~R. 2010, \aap, 510, A46.

\bibitem[\protect\astroncite{{Le Bars} \emph{et~al.}}{2015}]{lebars15}
{Le Bars}, M., {Lecoanet}, D., {Perrard}, S., {Ribeiro}, A., {Rodet}, L.,
  \emph{et~al.} 2015, Fluid Dynamics Research, 47, 045502.

\bibitem[\protect\astroncite{{Lecoanet} \emph{et~al.}}{2015}]{lecoanet15}
{Lecoanet}, D., {Le Bars}, M., {Burns}, K.~J., {Vasil}, G.~M., {Brown}, B.~P.,
  \emph{et~al.} 2015, \pre, 91, 063016.

\bibitem[\protect\astroncite{{Lecoanet} \& {Quataert}}{2013}]{lecoanet13}
{Lecoanet}, D. \& {Quataert}, E. 2013, \mnras, 430, 2363.

\bibitem[\protect\astroncite{{Lecoanet} \emph{et~al.}}{2017}]{lecoanet17}
{Lecoanet}, D., {Vasil}, G.~M., {Fuller}, J., {Cantiello}, M., \& {Burns},
  K.~J. 2017, \mnras, 466, 2181.

\bibitem[\protect\astroncite{{Lee} \& {Saio}}{1997}]{lee97}
{Lee}, U. \& {Saio}, H. 1997, \apj, 491, 839.

\bibitem[\protect\astroncite{{Malkus}}{1954}]{malkus54}
{Malkus}, W.~V.~R. 1954, Proceedings of the Royal Society of London Series A,
  225, 196.

\bibitem[\protect\astroncite{{Mathis}}{2009}]{mathis09}
{Mathis}, S. 2009, \aap, 506, 811.

\bibitem[\protect\astroncite{{Mathis}}{2013}]{mathis13a}
{Mathis}, S. 2013, In \emph{Lecture Notes in Physics, Berlin Springer Verlag},
  edited by M.~{Goupil}, K.~{Belkacem}, C.~{Neiner}, F.~{Ligni{\`e}res}, \&
  J.~J. {Green}, \emph{Lecture Notes in Physics, Berlin Springer Verlag}, vol.
  865, p.~23.

\bibitem[\protect\astroncite{{Mathis} \emph{et~al.}}{2016}]{mathis16}
{Mathis}, S., {Auclair-Desrotour}, P., {Guenel}, M., {Gallet}, F., \& {Le
  Poncin-Lafitte}, C. 2016, \aap, 592, A33.

\bibitem[\protect\astroncite{{Mathis} \emph{et~al.}}{2014}]{mathis14}
{Mathis}, S., {Neiner}, C., \& {Tran Minh}, N. 2014, \aap, 565, A47.

\bibitem[\protect\astroncite{{Mathis} \emph{et~al.}}{2008}]{mathis08}
{Mathis}, S., {Talon}, S., {Pantillon}, F.-P., \& {Zahn}, J.-P. 2008, \solphys,
  251, 101.

\bibitem[\protect\astroncite{{Miesch} \& {Toomre}}{2009}]{miesch09}
{Miesch}, M.~S. \& {Toomre}, J. 2009, Annual Review of Fluid Mechanics, 41,
  317.

\bibitem[\protect\astroncite{{Neiner} \emph{et~al.}}{2012}]{neiner12a}
{Neiner}, C., {Mathis}, S., {Saio}, H., {Lovekin}, C., {Eggenberger}, P.,
  \emph{et~al.} 2012, \aap, 539, A90.

\bibitem[\protect\astroncite{{Pin{\c c}on} \emph{et~al.}}{2016}]{pincon16}
{Pin{\c c}on}, C., {Belkacem}, K., \& {Goupil}, M.~J. 2016, \aap, 588, A122.

\bibitem[\protect\astroncite{{Pin{\c c}on} \emph{et~al.}}{2017}]{pincon17}
{Pin{\c c}on}, C., {Belkacem}, K., \& {Goupil}, M.-J. 2017, In \emph{European
  Physical Journal Web of Conferences}, \emph{European Physical Journal Web of
  Conferences}, vol. 160, p. 02002.

\bibitem[\protect\astroncite{{Prat} \emph{et~al.}}{2016}]{prat16}
{Prat}, V., {Ligni{\`e}res}, F., \& {Ballot}, J. 2016, \aap, 587, A110.

\bibitem[\protect\astroncite{{Press}}{1981}]{press81}
{Press}, W.~H. 1981, \apj, 245, 286.

\bibitem[\protect\astroncite{{Rogers}}{2015}]{rogers15}
{Rogers}, T.~M. 2015, \apjl, 815, L30.

\bibitem[\protect\astroncite{{Rogers} \emph{et~al.}}{2013}]{rogers13}
{Rogers}, T.~M., {Lin}, D.~N.~C., {McElwaine}, J.~N., \& {Lau}, H.~H.~B. 2013,
  \apj, 772, 21.

\bibitem[\protect\astroncite{{Rogers} \& {McElwaine}}{2017}]{rogers17}
{Rogers}, T.~M. \& {McElwaine}, J.~N. 2017, \apjl, 848, L1.

\bibitem[\protect\astroncite{{Saio} \emph{et~al.}}{2015}]{saio15}
{Saio}, H., {Kurtz}, D.~W., {Takata}, M., {Shibahashi}, H., {Murphy}, S.~J.,
  \emph{et~al.} 2015, \mnras, 447, 3264.

\bibitem[\protect\astroncite{{Samadi} \emph{et~al.}}{2010}]{samadi10}
{Samadi}, R., {Belkacem}, K., {Goupil}, M.~J., {Dupret}, M.-A., {Brun}, A.~S.,
  \emph{et~al.} 2010, \apss, 328, 253.

\bibitem[\protect\astroncite{{Schatzman}}{1993}]{schatzman93}
{Schatzman}, E. 1993, \aap, 279, 431.

\bibitem[\protect\astroncite{{Stello} \emph{et~al.}}{2016}]{stello16}
{Stello}, D., {Cantiello}, M., {Fuller}, J., {Huber}, D., {Garc{\'{\i}}a},
  R.~A., \emph{et~al.} 2016, \nat, 529, 364.

\bibitem[\protect\astroncite{{Stevenson}}{1979}]{stevenson79}
{Stevenson}, D.~J. 1979, Geophysical and Astrophysical Fluid Dynamics, 12, 139.

\bibitem[\protect\astroncite{{Talon} \& {Charbonnel}}{2005}]{talon05}
{Talon}, S. \& {Charbonnel}, C. 2005, \aap, 440, 981.

\bibitem[\protect\astroncite{{Talon} \& {Charbonnel}}{2008}]{talon08}
{Talon}, S. \& {Charbonnel}, C. 2008, \aap, 482, 597.

\bibitem[\protect\astroncite{{Zahn} \emph{et~al.}}{1997}]{zahn97}
{Zahn}, J.-P., {Talon}, S., \& {Matias}, J. 1997, \aap, 322, 320.

\end{thebibliography}

\end{document}